\documentclass[fleqn,twoside,epsf]{article}
\usepackage{espcrc2}

\usepackage{graphicx}

\newcommand{\AmS}{{\protect\the\textfont2
  A\kern-.1667em\lower.5ex\hbox{M}\kern-.125emS}}

\newcommand{\be}{\begin{equation}}
\newcommand{\ee}{\end{equation}}
\newcommand{\beq}{\begin{eqnarray}}
\newcommand{\eeq}{\end{eqnarray}}

\title{$\gamma N \rightarrow \Delta$ 
transition form factors 
in Quenched and $N_F=2$ QCD\thanks{Talk presented by A.~Tsapalis}}

\author{C.~Alexandrou\address[Cyprus]{Department of Physics,
University of Cyprus, CY-1678 Nicosia, Cyprus},
Ph.\ de Forcrand\address{ETH-Z\"urich, CH-8093 Z\"urich and CERN Theory Division, CH-1211 Geneva 23, Switzerland},
Th.~Lippert\address[Wuppertal]{Department of Physics, University of
Wuppertal, D-42097 Wuppertal, Germany},
H.~Neff\thanks{Acknowledges funding from
 the European network ESOP (HPRN-CT-2000-00130) and  the
University of Cyprus}\address{Institute of Accelerating Systems and Applications and Department of Physics, University of Athens, Athens, Greece, and Physics Department, Boston 
University, Boston,  Massachusetts 02215, USA},
J.~W.~Negele\address[MIT]{Center for Theoretical Physics, Laboratory for
Nuclear Science and Department of Physics, Massachusetts Institute of
Technology, Cambridge, Massachusetts 02139, USA},
K.~Schilling\addressmark[Wuppertal],
W. Schroers\thanks{Supported by the Alexander 
von Humboldt Foundation}\addressmark[MIT]
and
A.~Tsapalis\thanks{Supported by the Levendis Foundation}\addressmark[Cyprus]
}

\begin{document}
 
\begin{abstract}
Calculations of the magnetic dipole, electric quadrupole and Coulomb quadrupole
amplitudes for the transition $\gamma N\rightarrow \Delta$ are presented
both in quenched QCD  and with two flavours of
degenerate dynamical quarks.
\vspace{1pc}
\end{abstract}

\maketitle

\vspace*{-2.cm}

\section{Introduction}

The  shape of the proton is a fundamental issue
in hadron structure that depends on QCD dynamics. The purpose
of this study is to gain insight as to what  the
mechanism  that produces the deformation is.
Deformation can be determined from hadron wave functions
obtained via two- and three- density gauge invariant correlators~\cite{AFT}.
However the intrinsic  quadrupole moment is not always 
observable  
and it vanishes  between hadronic states of spin less than one.
Therefore to determine the deformation of the proton
one searches for  quadrupole strength in the transition 
$\gamma N  \rightarrow \Delta(1232)$
with real or virtual photons. Spin-parity
selection rules allow a magnetic
 dipole, an electric quadrupole and a Coulomb
quadrupole amplitude, 
which in the Sachs~decomposition~are~given~in~terms~of~the~form factors
${\cal G}_{M1}$,  ${\cal G}_{E2}$ and  ${\cal G}_{C2}$ 
that depend on the momentum
transfer $q^2 = (p'-p)^2$. 
Deviation from zero of
the ratios~\cite{Leinweber,All}

\vspace*{-0.2cm}

\be
R_{EM} = -\;\frac{{\cal G}_{E2}}{{\cal G}_{M1}} \;\;\;\;\;\;\;\;
R_{SM} = -\;\frac{|{\bf q}|}{2m_\Delta}\;\frac{{\cal G}_{C2}}{{\cal G}_{M1}} 
\label{EMR}
\ee

\vspace*{-0.65cm}
\noindent
 indicates deformation of the nucleon and/or $\Delta$. 
These  ratios have been  accurately  measured in recent  electroproduction experiments
 at Bates and Jefferson Labs~\cite{Bates} at various
momentum transfers.

\vspace*{-0.3cm}

\section{Lattice matrix elements}

\vspace*{-0.3cm}

\begin{figure}[h]
\vspace*{-0.7cm}
\mbox{\includegraphics[height=3cm,width=5cm]{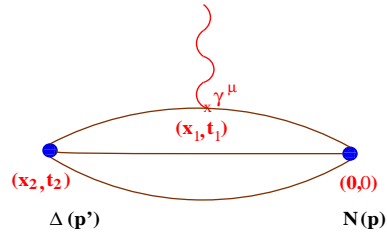}}
\vspace*{-0.8cm}
\caption{$\gamma N \rightarrow \Delta$ matrix element. The photon couples
to a quark in the nucleon
  at time separation $t_1$ from the source to produce a $\Delta$.}
\label{NDelta}
\vspace*{-0.8cm}
\end{figure}

We evaluate the 3-point function,
$\langle G^{\Delta j^\mu N}_{\sigma} (t ;{\bf p}^{\;\prime}, {\bf p} ; \Gamma) \rangle$
shown schematically in Fig.~\ref{NDelta}, as well as 
 $\langle G^{N j^\mu \Delta}_{\sigma}
(t_2, t_1 ; {\bf p}^{\;\prime}, {\bf p}; \Gamma) \rangle$ for the
reverse $\Delta \rightarrow \gamma N$ transition.
Exponential decays and normalization constants cancel in the ratio
$R_\sigma (t_2, t_1; {\bf p}^{\; \prime}, {\bf p}\; ; \Gamma ; \mu)=$

\vspace{-0.4cm}

\beq
& \>&\hspace*{-1.2cm}\large{
 \left [\frac{
\langle G^{\Delta j^\mu N}_{\sigma} (t_2, t_1 ; {\bf p}^{\;\prime}, {\bf p};
\Gamma ) \rangle \;
\langle G^{N j^\mu \Delta}_{\sigma} (t_2, t_1 ; -{\bf p}, -{\bf p}^{\;\prime};
\Gamma^\dagger ) \rangle }
{
\langle \delta_{ij} G^{\Delta \Delta}_{ij}(t_2,{\bf p}^{\; \prime};
\Gamma_4) \rangle \;
\langle G^{NN} (t_2, -{\bf p} ; \Gamma_4) \rangle } \right]^{1/2}
} \nonumber \\
&\;&\hspace*{-.5cm}\stackrel{t_2 -t_1 \gg 1, t_1 \gg 1}{\Rightarrow}
\Pi_{\sigma}({\bf p}^{\; \prime}, {\bf p}\; ; \Gamma ; \mu) \; ,
\hspace*{2cm} (2) \nonumber 
\label{R-ratio}
\eeq
where $ G^{NN}$ and $ G^{\Delta \Delta}_{ij}$ are the
 nucleon and $\Delta$ two point functions evaluated in the standard way.

\stepcounter{equation}

\noindent
The Sachs form factors are obtained by appropriate combinations
of the $\Delta$ spin-index $\sigma$, current direction $\mu$
and projection matrices $\Gamma$. For instance,
in the
$\Delta$ rest frame $ \vec{p}^{\;\prime} = 0 \;, \; \vec{q} = (q,0,0)$
~\cite{Leinweber,All} we have

\vspace*{-0.3cm}

$$
 \hspace*{-2.5cm}{\cal G}_{M1} = {\cal A}\;\frac{1}{|{\bf q} |} \;
\Pi_2({\bf 0}, {\bf -q}\; ; +i \;\Gamma_4 ; 3) 
$$

\vspace*{-0.5cm}



$$
{\cal G}_{E2} = {\cal A} \; \frac{1}{3 |{\bf q} |} \;
\Biggl[ \Pi_3({\bf 0}, {\bf -q}\; ; \Gamma_1 ; 3)  +
\Pi_1( {\bf 0}, {\bf -q}\; ; \Gamma_3 ; 3)  \Biggl] 
$$

\vspace*{-0.3cm}
\be
{\cal G}_{C2} = {\cal A} \; \frac{M_\Delta}{{\bf q}^{\;2}} \;
\Pi_1({\bf 0}, {\bf -q}\; ; -i \;\Gamma_1 ; 4) 
\ee

\noindent
where ${\cal A}$ is a kinematical factor.

\begin{figure}[h]
\vspace*{-0.9cm}
\tabskip=0pt\halign to\hsize{\hfil#\hfil\tabskip=0pt plus1em&
\hfil#\hfil\cr
\mbox{\includegraphics[height=2.6cm,width=3.5cm]{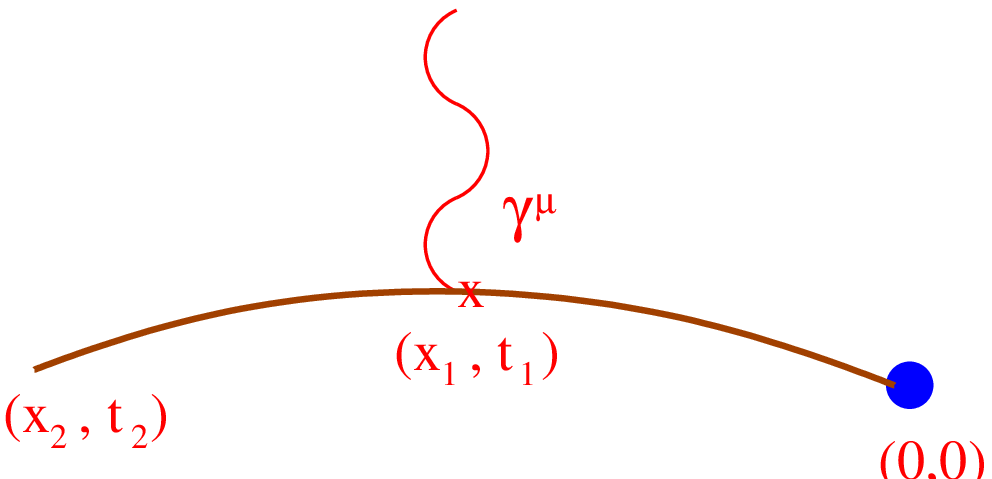}}&
\mbox{\includegraphics[height=2.6cm,width=3.5cm,]{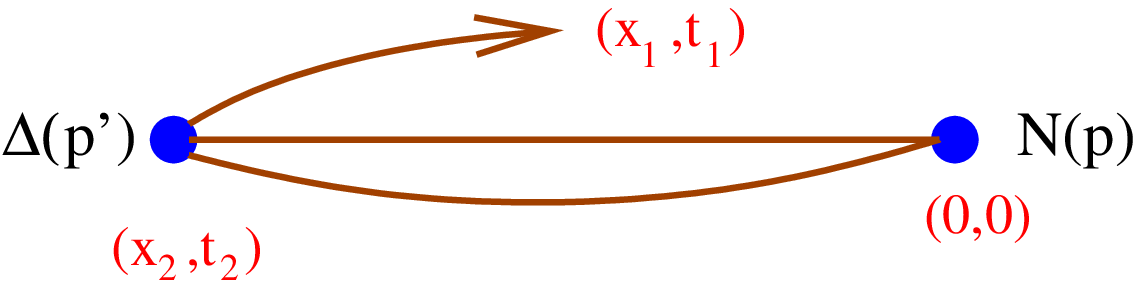}}\cr
(a)&(b)\cr}
\vspace*{-0.75cm}
\caption{(a) Fixed operator and (b) fixed source sequential propagators.}
\label{sequential}
\vspace*{-0.75cm}
\end{figure}

We use 
 two methods
 to compute the sequential propagator needed to build the 3-point function:
({\it a}) We evaluate the quark line with the photon insertion, 
shown schematically in Fig.~\ref{sequential}(a) by computing
the sequential propagator at fixed momentum transfer ${\bf q}$
and fixed time $t_1$. We look for a plateau by varying the sink-source
 separation
time  $t_2$. The final and initial states can be chosen at
the end.
({\it b}) We evaluate the backward sequential propagator shown schematically
in Fig.~\ref{sequential}(b) by fixing the initial and final states.
$t_2$ is fixed and a plateau is searched for by varying $t_1$. Since the
momentum transfer is specified only at the end the $\gamma N\rightarrow \Delta$
form factors can
be evaluated at all lattice momenta.

\begin{table}
\caption{}
\vspace*{-0.3cm}
\small
\label{table:parameters}
\begin{tabular}{|c|c|c|c|} \hline
 \multicolumn{1}{|c|}{$ Q^2$ (GeV$^2$)} &$\kappa$ & $m_\pi/m_\rho$ & Number of confs \\ \hline
\multicolumn{4}{|c|} {Quenched $\beta=6.0$ $16^3\times 32$ ${\bf q}^2 =0.64$ GeV$^2$} \\ \hline  
0.64   & 0.1530 & 0.84 & 100 \\ 
0.64   & 0.1540 & 0.78 & 100 \\
0.64   & 0.1550 & 0.70 & 100 \\ \hline 
\multicolumn{4}{|c|} {Quenched $\beta=6.0$ $32^3\times 64$ $ {\bf q}^2=0.64$ GeV$^2$ } \\ \hline
0.64   & 0.1550 & 0.69 & 100 \\ \hline
 \multicolumn{1}{|c|}{$ Q^2$ (GeV$^2$)} &$\kappa$ & $m_\pi/m_\rho$ & Number of confs \\  \hline
0.16  & 0.1554 & 0.64 & 100 \\  
0.15  & 0.1558 & 0.59 & 100  \\    
0.13  & 0.1562 & 0.50 & 100   \\ \hline
\multicolumn{4}{|c|} {Unquenched $\beta=5.6$ $16^3\times 32$~\cite{SESAM} $ {\bf q}^2=0.54$ GeV$^2$ }\\ \hline
0.54   & 0.1560  & 0.83 & 196\\  
0.54   & 0.1565  & 0.81 & 200  \\
0.54   & 0.1570  & 0.76 & 201  \\
0.54   & 0.1575  & 0.68 & 200  \\ \hline
\end{tabular}
\vspace*{-1.cm}
\end{table}

\vspace*{-0.2cm}

\section{Results}

\vspace*{-0.2cm}

Using the fixed operator method for the sequential propagator  
we  obtain, for the same value of ${\bf q}$,
 two kinematically different cases:
one with  the $\Delta$  at rest and the other with the nucleon
at rest. 
The parameters of our lattices are given in Table~\ref{table:parameters} where
 we used  the nucleon  mass in the chiral
limit to convert to physical units, and $Q^2=-q^2$ is evaluated in the rest frame of the $
\Delta$.

\begin{figure}[h]
\vspace*{-0.5cm}
\tabskip=0pt\halign to\hsize{\hfil#\hfil\tabskip=0pt plus1em&
\hfil#\hfil&\hfil#\hfil\tabskip=0pt\cr
\mbox{\includegraphics[height=5cm,width=6cm]{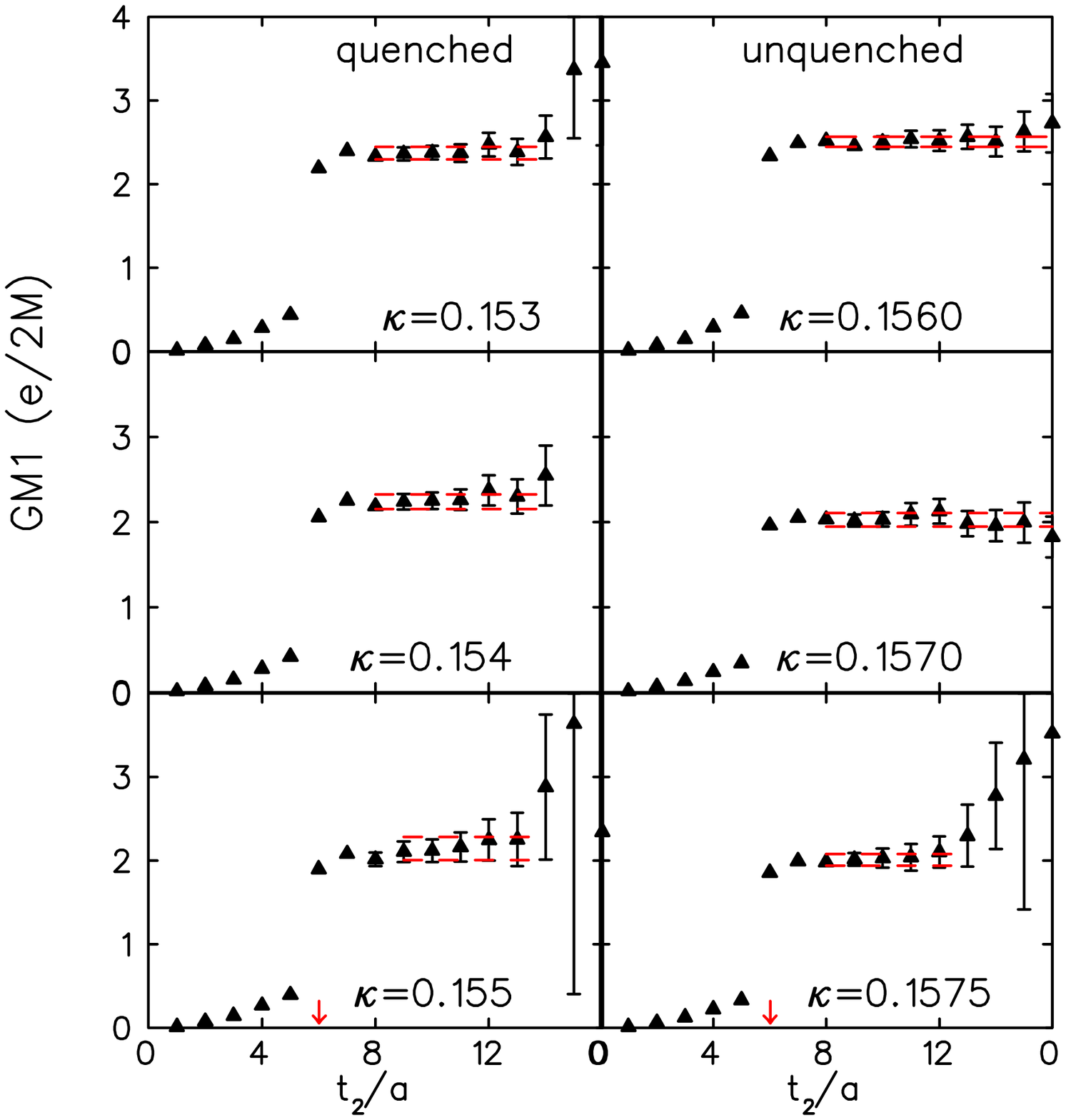}}& \hspace*{0.1cm}
\mbox{\includegraphics[height=5cm,width=6cm]{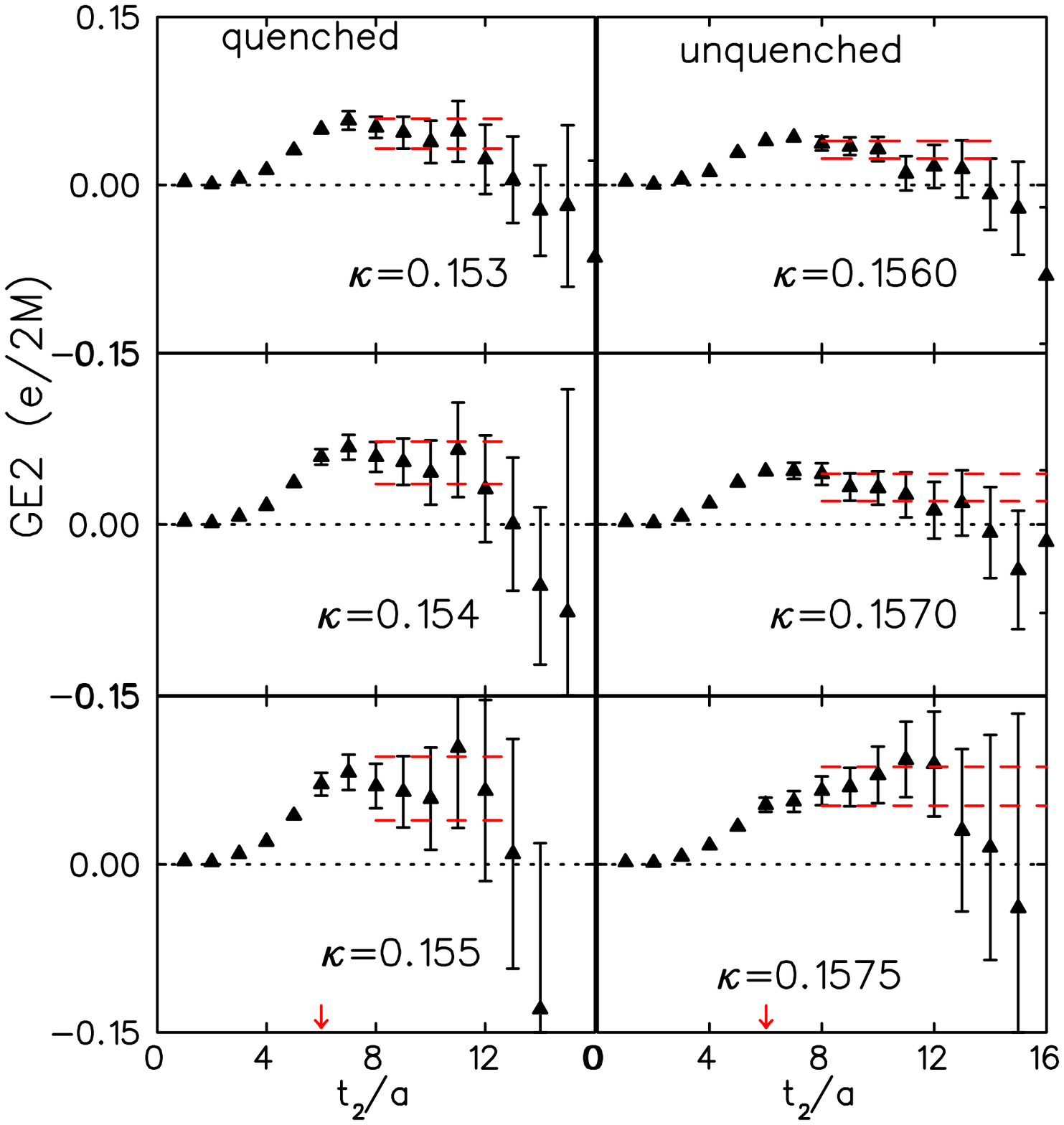}}&
\hspace*{0.1cm}\mbox{\includegraphics[height=5cm,width=4cm]{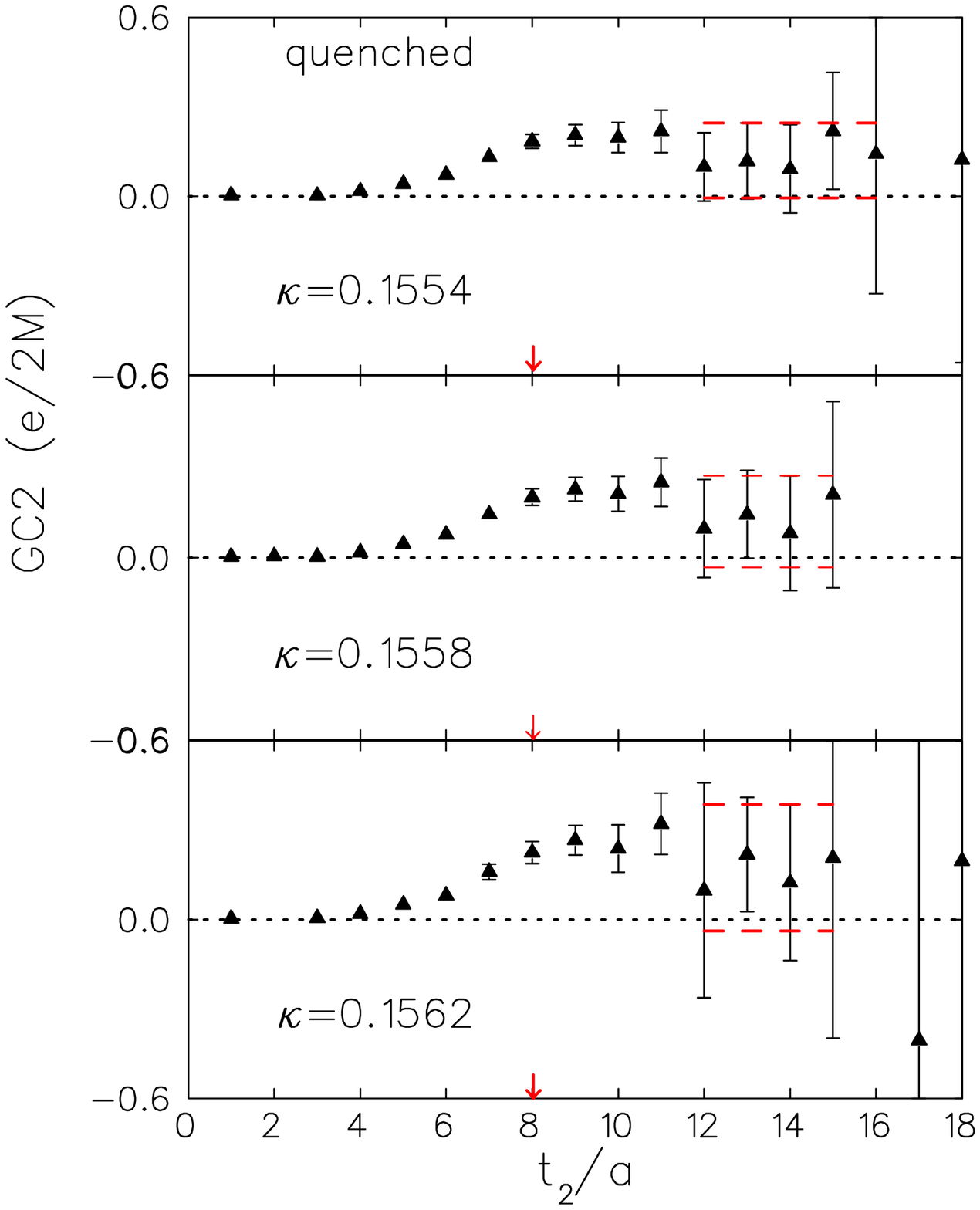}}\cr
${\cal G}_{M1}$ at  ${\bf q}=(2\pi/16a,0,0)$,  $t_1=6a$ &
${\cal G}_{E2}$ at ${\bf q}=(2\pi/16a,0,0)$,  $t_1=6a$ &${\cal G}_{C2}$ at ${\bf q}=(2\pi/32a,0,0)$,  $t_1=8a$\cr}
\vspace*{-0.8cm}
\caption{
The dashed 
lines show the plateau fit range and 
bounds obtained by jackknife analysis.}
\label{fig:all}
\vspace*{-0.7cm}
\end{figure}

 We check for finite volume effects
by comparing results
in the quenched theory on
lattices of size $16^3\times 32$ and $32^3\times 64$ 
at  the same momentum transfer at $\kappa=0.1550$. 
Assuming a 1/volume dependence we find that on the small
volumes there is a $(10-15)\%$ correction as compared to the infinite
volume result whereas 
on the large lattice the volume correction  is negligible.
In Fig.~\ref{fig:all} we  show quenched and
unquenched results for ${\cal G}_{M1}$ and ${\cal G}_{E2}$ at the same momentum transfer and similar ratios of pion to rho mass.
Unquenching 
leads to a stronger mass dependence but 
leaves
 the ratio $R_{EM}$ largely unaffected, giving values in the range of $-(2-4)\%$.
 The fact that no
increase of  $R_{EM}$ is observed
means that pion contributions to this ratio are small for these heavy pions.
In Fig.~\ref{fig:all} we also show results for
 ${\cal G}_{C2}$ with $\Delta$ static
for the large quenched lattice for which
we obtain the best signal.
Although  ${\cal G}_{C2}$ is within one standard deviation of zero, it is positive
at all $\kappa$-values giving  a negative 
$R_{SM}$ in agreement with experiment.


\begin{figure}[h]
\vspace*{-0.8cm}
\mbox{\includegraphics[height=4.9cm,width=7cm]{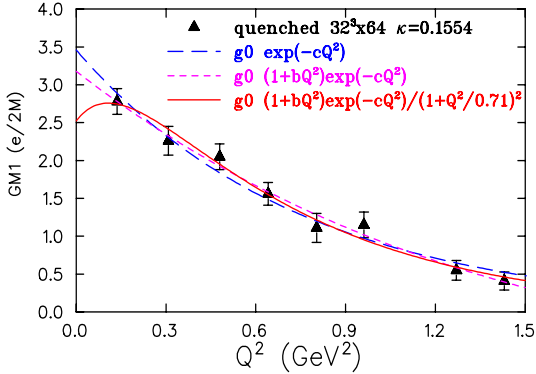}}
\vspace*{-1cm}
\caption{${\cal G}_{M1}$ versus $Q^2$ 
  in the  $\Delta$ rest frame.}
\vspace*{-0.8cm}
\label{GM1-q2}
\end{figure}

Using the fixed sink method  for the sequential 
propagator we show in Fig.~\ref{GM1-q2}
 preliminary results for  the $q^2$-dependence of
${\cal G}_{M1}$
obtained using 50 quenched configurations  at $\kappa=0.1554$ together
with a fit to the preferred phenomenological form
$g_0(1+bQ^2)\exp(-cQ^2)F_D$ where $F_D=1/(1+Q^2/0.71)^2$ is the nucleon dipole
form factor.  Although lattice data
are consistent 
with this form other parameterizations, such as simple exponential 
dependence, can not presently be excluded. 

\begin{table}
\vspace*{5.5cm}
\caption{}
\label{table:chiral}
\small
\vspace*{-0.2cm}
\begin{tabular}{|c|c|c|c|} 
\hline
 $Q^2$~GeV$^2$ &  ${\cal G}_{M1}\>(e/2m_N)$ &  ${\cal G}_{E2}\>(e/2m_N)$ & {$R_{EM}\%$} \\ \hline
\multicolumn{4}{|c|}{Quenched QCD} \\ \hline
  0.64 &  1.70(6)  & 0.103(19) & -5.1(1.1) \\
 0.13 &   2.53(6) & 0.103(12) & -4.5(1.4) \\\hline
\multicolumn{4}{|c|}{Unquenched QCD}  \\ \hline
  0.53 &  1.30(4)  & 0.052(31) & -2.9(1.5) \\ 
\hline
\end{tabular}
\vspace*{-0.8cm}
\end{table}

Chiral extrapolation of the results is done  
linearly in the pion mass squared,
since with the nucleon or the $\Delta$ carrying
a finite momentum,  chiral logs are expected to be suppressed.
The values obtained are given in  Table~\ref{table:chiral}
and are in reasonable agreement with the experimental values 
$R_{EM}=-2.1 \pm 0.2 \pm 2.0$ at  $Q^2=0.126$~GeV$^2$ and 
$R_{EM}=-1.6 \pm 0.4 \pm 0.4$  at  $Q^2=0.52$~GeV$^2$.

\section{Conclusions}

1)  The ratio $R_{EM}$  has been computed for the first time
with enough accuracy to exclude a zero value. In the
kinematical regime explored by experiments we obtain values which are
in agreement with recent measurements. 
2) Large statistical and systematic errors prevent an accurate
determination of the Coulomb quadrupole form factor. Although a zero
value cannot be excluded, our results support a negative value of 
the ratio $R_{SM}$ in agreement with experiment.
3) The detailed $Q^2$ dependence of the magnetic dipole transition 
can be evaluated with $\sim$ 10\% accuracy in the regime explored by JLab.
4) For pions in the range of 800-500 MeV no unquenching effects can be
established for $R_{EM}$ within our statistics.

\end{document}